# Axial cosinoidal structured illumination using phase modulation

Xianlin Song [a, #, *], Jianshuang Wei [b, c, #], Lingfang Song [d]
[a] School of Information Engineering, Nanchang University, Nanchang 330031, China;
[b] Britton Chance Center for Biomedical Photonics, Wuhan National Laboratory for Optoelectronics-Huazhong University of Science and Technology, Wuhan 430074, China;
[c] Moe Key Laboratory of Biomedical Photonics of Ministry of Education, Department of Biomedical Engineering, Huazhong University of Science and Technology, Wuhan 430074, China;
[d] Nanchang Normal University, Nanchang 330031, China;
# equally contributed to this work
* songxianlin@ncu.edu.cn

## ABSTRACT

We propose a novel method for generating axial cosine structured light by using phase-only spatial light modulator. We implemented axial cosine structured light using holographic technique. The computer generates double concentric annular slits with different radius, a prism phase is applied on the slits to tilts the beam that incidents on the slits away from the optical axis. Different annular beam produce Bessel beams with different axial wave vectors, axial cosine structured light can be obtained from the interference between two Bessel beams. The period and phase of axial cosine structured light can be adjusted by adjusting the radius and the initial phase difference of the double concentric annular beams. We theoretically and experimentally verify that the method can effectively generate axial cosine structured light.

## 1. INTRODUCTION

Axial cosine structured light is a beam that the distribution of the axial intensity is cosine distribution. It has been widely used in many fields, such as optical tweezers[1-2], which can control the micron particles, nanoparticles, free electrons, biological cells and atoms or molecules; In structured illumination microscopy[3-4], special devices such as digital micromirror device (DMD) are used to generate cosine structured light to modulate signals, the high frequency information is moved to the low frequency detection area due to the Moire effect. The quality of cosine structured light directly determines the improvement of resolution. At present, there are many ways to generate axial cosine structured light. A mask plate with double concentric annular slits with different radius is used to generate double concentric annular beams[5]. When the mask plate is placed on the focal plane of the lens, each ring will generates Bessel beam with different axial wave vectors at the focal zone of the lens, and then interference generates axial cosine structural light. The two Bessel beams interfere to produce axial cosine structured light. This method requires the fabrication of a mask, and once the mask is completed, the phase and period of the structural light cannot be adjusted by this method. A double-negative axicon chemically etched in the optical fiber tips was used to generate axial cosine structured light[6]. The double-negative axicon can generate $\pi$ –phase shifted multi-ring hollow Gaussian beam with different axial wave vectors, and then each ring gerenates non-diffracting Bessel beams resulting in the generation of cosinoidal intensity distribution along the propagation direction. However, the implementation of this method is complicated, it is difficult to regulate the period and phase of the generated structured light. Since axicon is a very common device to produce zero order Bessel beam, Two axicons with different base angle are also used to generate axial cosine structured light[7,8]. Tuanjie Du used an axicon to generate non-diffracting Bessel beam, then the non-diffracting Bessel beam was focused by the second axicon with different base angle, thus the axial cosine structured light was generated[7]. Fengtie Wu used two axicons with different base angle to generate Bessel beams with different axial wave vectors, then, the two Bessel beams interfere to produce axial cosine structured light[8]. Although the method of using axicon is simpler, the period and phase of axial cosine structured light are fixed and cannot be changed.

In this manuscript, we report a novel method to address the problem, we employ holographic technique to produce double concentric annular beams. The computer generates double concentric annular slits with different radius, a prism phase is applied on the slits to tilts the beam that incidents on the slits away from the optical axis. Different annular beam produce Bessel beams with different axial wave vectors, axial cosine structured light can be obtained from the interference between two Bessel beams. The period and phase of axial cosine structured light can be adjusted by adjusting the radius and the initial phase difference of the double concentric annular beams. We theoretically and experimentally verify that the method can effectively generate axial cosine structured light.

## 2. METHOD

### 2.1 Principle

Bessel beam is a non - diffracted beam, it has self-reconstructing characters that can continue to spread and return to its original state when it encounters obstacles. One of the common way to generate Bessel beams is to use an annular slit, placing the annular slit on the back focal plane of the lens[9]. The radius of the annular slit determines the axial wave vector of the Bessel beams, and the width of the annular slit determines the intensity. Therefore, some researchers use concentric double annular slits to generate two Bessel beams with different axial wave vectors, and their interference on the axis will generate periodic intensity distribution[10]. A concentric double annular slits with inner radius $r_1$ and outer radius $r_2$ is placed in front of the lens with the focal length $f$. In polar coordinate system, the complex amplitude of the light field $U_1(r,z)$ passing through the double annular slits is

$$U_1(r) = [circ(\frac{r}{r_1 + \Delta r_1}) - circ(\frac{r}{r_1})] + [circ(\frac{r}{r_2 + \Delta r_2}) - circ(\frac{r}{r_2})] \tag{1}$$

While, $circ(r/r_1)$ and $circ(r/r_2)$ represent circular function of radius $r_1$ and $r_2$, respectively. $\Delta r_1$ and $\Delta r_2$ are the silt widths of the inner and the outer annular slit, respectively. The diffraction field on the output plane of z after the double concentric annular beams passed through the lens can be approximately expressed as ($\Delta r_1$ and $\Delta r_2 \to 0$, which means the silts are very narrow)

$$U(r,z) = C2\pi r_1 \Delta r_1 \exp(\frac{j\pi}{\lambda} ar^2) \exp(\frac{j\pi}{\lambda} br_1^2) J_0\left[\frac{2\pi r_1 rf}{\lambda(fz - z_0 z + z_0 f)}\right]$$
$$+ C2\pi r_2 \Delta r_2 \exp(\frac{j\pi}{\lambda} ar^2) \exp(\frac{j\pi}{\lambda} br_2^2 + j\Delta\phi) J_0\left[\frac{2\pi r_2 rf}{\lambda(fz - z_0 z + z_0 f)}\right] \tag{2}$$

and

$$a = \frac{-z_0 f}{(fz - z_0 z + z_0 f)z} + \frac{1}{z}$$
$$b = \frac{-zf}{(fz - z_0 z + z_0 f)z} + \frac{1}{z_0} \tag{3}$$

While, $C$ is integral constant, $z_0$ is the distance between the slit and the lens, $J_0$ is the zero order Bessel function. The superimposed field is formed by the superposition of two Bessel beams generated by two annular slits. When the annular slits is placed on the back focal plane of the lens ($z_0 = f$), Equation (2) can be written as

$$U(r,z) = r_1 \Delta r_1 J_0\left[\frac{2\pi r_1 r}{\lambda f}\right] \exp(\frac{j\pi}{\lambda} \frac{f-z}{f^2} r_1^2)$$
$$+ r_2 \Delta r_2 J_0\left[\frac{2\pi r_2 r}{\lambda f}\right] \exp(\frac{j\pi}{\lambda} \frac{f-z}{f^2} r_2^2 + j\Delta\phi) \tag{4}$$

Spatial intensity distribution of total diffracted field is

$$I(r,z) = |U(r,z)|^2 = r_1^2 \Delta r_1^2 J_0^2(\frac{2\pi r_1 r}{\lambda f}) + r_2^2 \Delta r_2^2 J_0^2(\frac{2\pi r_2 r}{\lambda f})$$
$$+ 2r_1 r_2 \Delta r_1 \Delta r_2 J_0(\frac{2\pi r_1 r}{\lambda f}) J_0(\frac{2\pi r_2 r}{\lambda f}) \cos(\Delta\theta) \quad (5)$$

and

$$\Delta\theta = \frac{\pi}{\lambda} \frac{f-z}{f^2}(r_1^2 - r_2^2) + \Delta\phi \quad (6)$$

While, $\Delta\phi = \phi_0$ is the initial phase difference of the double annular slits. In equation (5), the first two terms are constant terms. Since the two Bessel beams have different axial wave vectors, the phase difference between the Bessel beams is change with the propagation distance. Thus, the entire intensity distribution is shown as the axial cosine distribution. The period of structured light is calculated as Phase diagram generation

$$T = \frac{2\lambda f^2}{r_1^2 - r_2^2} \quad (7)$$

The period of the axial structured light can be changed by changing the radius of the inner and outer annular slits, and the phase shift of the structured light can be made by changing the initial phase difference of the double annular slits.

## 2.2 Generation of double annular beams using hologram

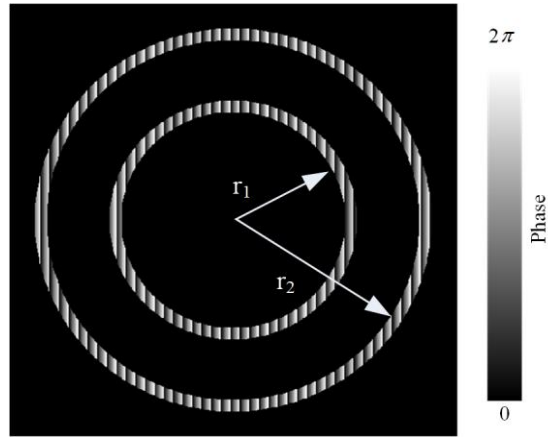

Figure 1. A phase diagram that produces double annular beams. The prism phase in the double annular slits deflects the beam from the optical axis.

As can be seen above, the regulation of the period and phase of axial cosine structured light is equivalent to the regulation of the radius and initial phase difference of the double annular beams. Here, holographic technique is used to produce double concentric annular beams with adjustable radius and initial phase difference. The hologram is shown in Figure 1, the phase map is contain double concentric annular silts. The inner annular silt has a radius of $r_1$ and a silt width of $\Delta r_1$, the outer annular silt has a radius of $r_2$ and a silt width of $\Delta r_2$. The prism phase was added into the silts, making the beam incident on the annular silts tilted from the optical axis. The beam that incidents on the other part of the phase map without prism phase continues to propagate along the optical axis. Thus, the double annular beams can be achieved.

### 2.3 System for generation of axial cosine structured light

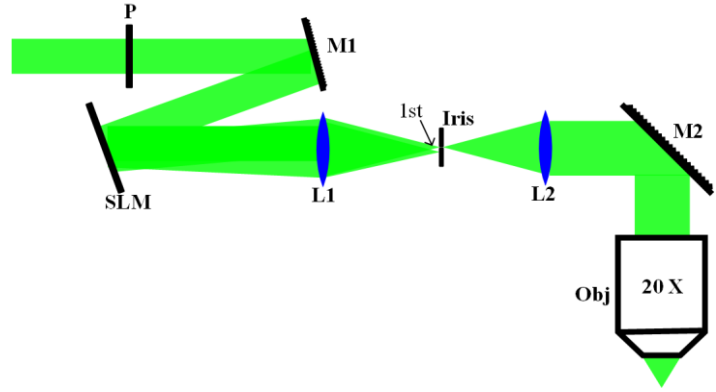

Figure 2. System for generation of axial cosine structured light. L1, L2, lens; M1, M2, mirror; Obj, objective; P, polarizer; SLM, phase only spatial light modulator.

The system for generation of axial cosine structured light is shown in Figure 2. The collimating laser beam (the wavelength is 523 nm) was horizontally polarized and obliquely directed onto the phase only spatial light modulator (SLM, Pluto VIS 006, Holoeye). The phase hologram is loaded into the SLM before the beam was modulated and focused by a concave lens L1 (f=180 mm). By placing an iris at the focus of the lens, only the first-order diffracted light (modulated light) is reserved. Then the beam is collimated by a concave lens L2 (f=180 mm). Finally, the beam focused by a water immersion objective (20×, N.A. = 1, Olympus). We obtain the axial structured light at the focal zone of the objective.

## 3. RESULTS

### 3.1 Simulation

To verify the validity of our method, a simulation was carried out according to the equation (5). The inner annular silt has a radius of 320 μm and the outer annular silt has a radius of 1600 μm. Figure 3 is the simulation result, Figure 3(a) is the axial cosine structured light, the intensity of the light shows a cosine distribution along the axis. The period is ~ 34 μm. At the same time, some weak side lobes are distributed around the axis, this may be caused by the superposition of side-lobe of the Bessel beams. Figure 3(b) is the cross-section view of the maximum intensity of the axial cosine structured light indicated by the white dashed line 1 in Figure 3(a). The Bessel beams produced by the two annular silts constructive interference with each other at this position, and the light intensity reached its maximum. Figure 3(c) is the cross-section view of the minimum intensity of the axial cosine structured light indicated by the white dashed line 2 in Figure 3(a). The Bessel beams produced by the two annular silts destructive interference with each other at this position, and the light intensity reached its minimum.

### 3.2 Experiment

A phase diagram that the inner annular silt has a radius of 320 μm and the outer annular silt has a radius of 1600 μm is loaded on the SLM. Figure 4 is the experiment result. Taking the position of the maximum light intensity as the reference point, we selected the spots at several locations in a period, as shown in Figure 4. Figure 4(a) is the cross-section view of the maximum intensity (z = 0 μm) of the axial cosine structured light. There is a bright spot in the center, and there are several weak side lobes around it. When the beam travels forward by 4.8 μm, the center spot become darken, as shown in Figure 4(b). When the beam travels forward by 16.8 μm, Due to the destructive interference of the two Bessel beams, the light intensity is minimized, a dark hole appears in the center, as shown in Figure 4(b). As the beam continues to travel, the intensity in the central gets stronger. When the beam travels forward by 33.6 μm, the intensity of the central spot is back to its strongest, the intensity distribution is the same as the reference point (z = 0 μm). Finally, the period of the axial cosine structured light is estimated to be 33.6 μm, which agree with theoretical prediction.

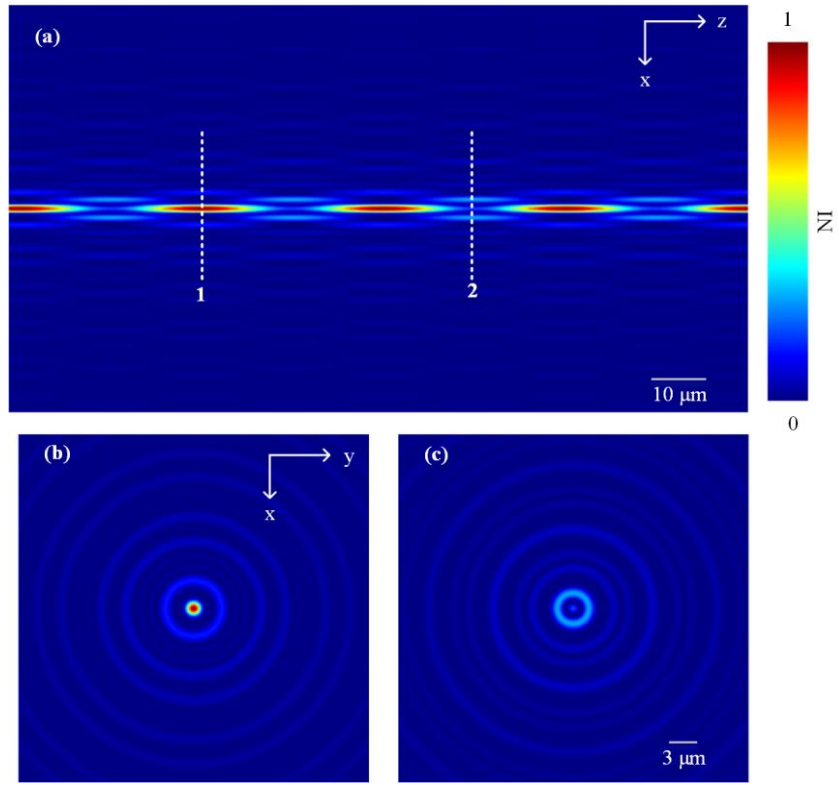

Figure 3. Simulation result of axial cosine structured light. NI, normalized intensity.

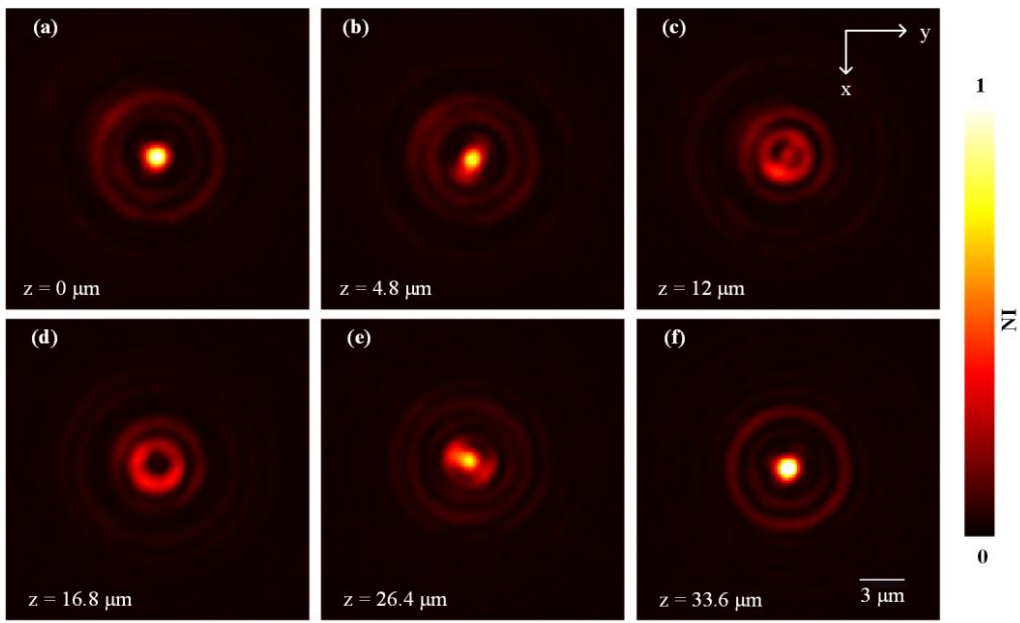

Figure 4. Experiment result of axial cosine structured light. NI, normalized intensity.

## 4. CONCLUSION

In summary, by using holographic technique, we developed a novel method for generating axial cosine structured light. The computer generates double concentric annular slits with different radius, a prism phase is applied on the slits to tilts the beam that incidents on the slits away from the optical axis. Different annular beam produce Bessel beams with different axial wave vectors, axial cosine structured light can be obtained from the interference between two Bessel beams. The period and phase of axial cosine structured light can be adjusted by adjusting the radius and the initial phase difference of the double concentric annular beams. Simulation and experiment are carried out to verify the effectiveness of the method, and the simulation results are in agreement with the experimental results. This method will be helpful to the development of technologies like optical tweezers, optical modulation.